\documentclass[]{aiaa-tc}

\usepackage{graphicx}
\usepackage{epstopdf}
\usepackage{amsmath}
\usepackage{amssymb}
\usepackage{amsfonts}
\usepackage{pifont}
\usepackage[small,bf]{caption}
\usepackage{subfig}
\usepackage{tikz}
\usetikzlibrary{shapes,arrows}
\usetikzlibrary{calc}
\usepackage{pgf}
\usepackage{pgffor}
\usepackage{subfig}
\usepackage{hyperref}
\usepackage{rotating}
\usepackage[sort&compress, numbers]{natbib}
 \usepackage{relsize}

\title{Performance limitations of observer-based feedback for transient energy growth suppression} 

 \author{
    Maziar S. Hemati%
    \thanks{\msh{Assistant Professor, Aerospace Engineering and Mechanics. Senior Member AIAA.}} 
     \ and Huaijin Yao%
         \thanks{\msh{Graduate Student, Aerospace Engineering and Mechanics. Member AIAA.}}\\
  {\normalsize\itshape
    \msh{University of Minnesota, Minneapolis, MN 55455}}
 }

 \AIAApapernumber{2017-}
 \AIAAconference{Conference Name, Date, and Location}
 \AIAAcopyright{\AIAAcopyrightD{Maziar S. Hemati}}

\newcommand{\diag}{\mathrm{diag}}
\newcommand{\trans}{^\mathsf{T}}

\renewcommand{\Re}{\mathbb{R}}

\newcommand{\id}{\mathrm{d}}

\newcommand{\ddt}{\frac{\mathrm{d}}{\mathrm{d}t}}
\newcommand{\cl}[1]{\tilde{#1}}
\newcommand{\mteg}{G} 
\newcommand{\mtegol}{G}
\newcommand{\mtegsp}{\cl{G}} 
\newcommand{\mtegspx}{\cl{G}_x}
\newcommand{\mtegspxhat}{\cl{G}_{\hat{x}}}

\newcommand{\Asp}{\cl{A}} 

\newcommand{\mteglemma}[1]{#1\trans+#1\le0}
\newcommand{\mlemma}{MTEG~Lemma}

\newcommand{\msh}[1]{{{\color{black}{#1}}}}

\usepackage{array}
\newcolumntype{L}[1]{>{\raggedright\let\newline\\\arraybackslash\hspace{0pt}}m{#1}}
\newcolumntype{C}[1]{>{\centering\let\newline\\\arraybackslash\hspace{0pt}}m{#1}}

\begin{document}

\maketitle

\begin{abstract}
Transient energy growth suppression is a common control objective for feedback 
flow control aimed at delaying transition to turbulence. 
A prevailing control approach in this context is observer-based feedback, 
in which a full-state feedback controller is applied to state estimates from an observer. 
The present study identifies a fundamental performance limitation of observer-based feedback control: 
whenever the uncontrolled system exhibits transient energy growth in response to optimal disturbances, control by observer-based feedback will necessarily lead to transient energy growth in response to optimal disturbances for the closed-loop system as well. 
Indeed, this result establishes that observer-based feedback can be a poor candidate for 
controller synthesis in the context of transient energy growth suppression and transition delay: 
the performance objective of transient energy growth suppression can never be achieved by means of observer-based feedback. 
%
%
Further, an illustrative example is used to show 
that alternative forms of output feedback are 
not necessarily subject to these 
same performance limitations, and should also be considered in the context of transient 
energy growth suppression and transition control.

\end{abstract}

\section*{Nomenclature}

\begin{tabular}{C{0.25\textwidth} L{0.75\textwidth}}
  $(A,B,C)$ & Linear time-invariant state-space realization of the plant\\
  $\Asp$ & Closed-loop observer-plant operator\\
  $E$ & Open-loop energy\\
  $\cl{E}$ & Closed-loop observer-plant energy\\
  $\mteg$ & Open-loop plant maximum transient energy growth\\
  $\mtegsp$ & Closed-loop observer-plant maximum transient energy growth\\
  $\mtegspx$ & Closed-loop plant maximum transient energy growth\\
  $\mtegspxhat$ & Closed-loop observer maximum transient energy growth\\
  $K$ & Controller feedback gain\\
  $L$ & Observer gain\\
  $u$, $x$, $y$ & Plant input, state, and output vectors, respectively\\
  $\hat{x}$ & Observer (estimated) state\\
  $\cl{x}=(x,\hat{x})$ & Closed-loop observer-plant state\\
\end{tabular}

\section{Introduction}
An ability to delay or fully suppress transition to turbulence has the potential to benefit a variety of technological systems, including air and maritime transportation systems, by enabling improvements to efficiency and performance.
Transition to turbulence in many shear flows  arises at 
a Reynolds number~($Re$) well below the critical $Re$ 
predicted by a linear stability analysis of 
the Navier-Stokes equations about a laminar solution~\cite{schmidBook}.
The onset of this so-called \emph{sub-critical transition} 
can often be explained by a linear mechanism for transient energy growth:
a non-modal stability analysis of the linearized dynamics reveals that, 
for linearly stable flows, small disturbances will often grow before 
they decay~\cite{trefethenScience1993,schmidAnnRev2007}.
Indeed, this linear mechanism for transient energy growth can cause 
the fluid state to depart from the basin of attraction for the laminar solution, 
triggering transition and ultimately giving rise to turbulence.

Numerous investigations have sought to delay transition 
by aiming to reduce transient energy growth through various forms of linear feedback control~\cite{joshiJFM1997,bewleyJFM1998,mckernan2007,monokrousos2010,martinelli2011,dadfar2013,bewley2001,kim2003,kimAnnRev2007,bagheri2009,bagheri2011}.
%
Many of these studies have relied upon the 
well-established separation principle at some stage in 
the synthesis of a dynamic output feedback compensator (i.e.,~observer-based feedback)~\cite{bewley2001,kim2003,kimAnnRev2007,bagheri2009,bagheri2011}.
The separation principle is commonly invoked in this design process because 
it considerably simplifies controller synthesis when a dynamic output feedback law is desired:
a stabilizing full-state feedback controller can be synthesized independently of a stable state estimator,
then combined together to yield a stabilizing dynamic output feedback compensator~\cite{brogan1991}.
Despite guarantees on linear stability of the closed-loop system,
invoking the separation principle can dramatically degrade closed-loop performance.
Yet, the adverse consequences of controller synthesis by means of 
the separation principle are not fully appreciated in the context 
of transition delay and transient energy growth control, in which 
closed-loop performance is paramount.
For instance, the separation principle is central to 
linear quadratic Gaussian~(LQG) control, which remains a 
common controller synthesis approach for transient 
energy growth reduction and transition delay~\cite{bewley2001,kim2003,kimAnnRev2007,bagheri2009,bagheri2011}.
Although degraded closed-loop performance of 
observer-based feedback controllers has been 
reported in the literature~\cite{bewleyJFM1998}, 
no previous studies have explicitly identified the separation 
principle as the source of these performance limitations.
\msh{The state estimation problem is sometimes identified to 
be ``the primary pacing item'' for realizing acceptable 
closed-loop performance~\cite{kimAnnRev2007};
however, as we will show here, \emph{observer peaking} and 
associated performance issues are more deeply rooted with 
the separation principle itself.}
%
%

%

In this study, 
we will show that any linear system that exhibits transient
energy growth to disturbances in open-loop will also exhibit transient energy growth to disturbances in closed-loop whenever the separation principle is invoked for controller synthesis.
Indeed, even when a full-state feedback controller can fully suppress transient energy growth,
a dynamic output feedback compensator designed via the separation principle will invariably lead to non-trivial transient energy growth in response to some disturbances.
Further, the separation principle will guarantee that the estimator 
dynamics (and estimation error) will exhibit peaking in closed-loop from some initial 
conditions whenever the uncontrolled plant exhibits 
transient energy growth.
%
%
The results in this study highlight the inherent performance limitations
that arise by invoking the separation principle for observer-based feedback control 
in the context of transition control and transient energy growth suppression.
%
%
As we will show via example, not all output feedback control approaches are restricted 
to the same performance limitations as observer-based feedback; 
indeed, alternative output feedback approaches should 
be considered as candidates for controller synthesis when closed-loop performance is 
a primary design objective.

\section{{Observer-based feedback and maximum transient energy growth}}
\label{sec:proofs}
Consider the linear time-invariant system,
\begin{equation}
\begin{split}
  \dot{x}(t) &= Ax(t)+Bu(t)\\
  y(t) &= Cx(t),
  \end{split}
  \label{eq:linsys}
\end{equation}
where $x\in\Re^n$ is the state vector, $u\in\Re^m$ is the input vector, 
$y\in\Re^p$ is the output vector, and $t\in\Re$ is time.
In considering transient energy growth in this study, 
we focus on the free response of the system to an initial perturbation $x(t_o)=x_o$ away from an equilibrium solution (e.g.,~a laminar base flow).
The perturbation dynamics are given in terms of the matrix exponential, 
$	x(t) = \mathrm{e}^{A(t-t_o)}x_o,$
%
and the associated perturbation energy will have a response given by,
\begin{equation}
{E(t) = x\trans(t) Qx(t)}, 
\label{eq:energy}
\end{equation}
where $Q=Q\trans>0$.
 Without loss of generality, we take $Q=I$, 
 since the state can always be transformed as~$\bar{x}=Q^{1/2}x$ 
 to satisfy this definition of energy.
Further, define the maximum transient energy growth $G$ as,
\begin{equation}
	\mteg = \max_{t\ge t_o}\max_{\substack{E(t_o)\ne0}} \frac{E(t)-E(t_o)}{E(t_o)} = \max_{t\ge t_o}\max_{\substack{\|x_o\|=1}} E(t)-1\ge0.
\end{equation}
When the system in~\eqref{eq:linsys} is unstable, $E(t)$ will be unbounded and $G$ will be infinite;
when the system in~\eqref{eq:linsys} is stable, $G$ is simply the peak value in 
transient energy growth due to a so-called \emph{optimal disturbance}~\cite{butlerPOF1992}.
%
%
The lower-bound $\mteg=0$ corresponds to the case of 
suppressed transient energy growth~(i.e.,~monotonic stability~\cite{schmidBook}), 
which is the ultimate aim of 
controllers designed for transient energy growth suppression.
%
%
%

In light of these definitions, we introduce a lemma~(herein, referred to 
as the \emph{\mlemma}) that will be central to the analysis here: $\mteg=0$ if and only if $\mteglemma{A}$~\cite{whidborne2007}.
%
%
%
%
To show sufficiency, consider that if $\dot{E}(t)\le0$ for all $t\ge t_o$ and all initial conditions, then $\mteg=0$.
Since $\dot{E}(t)=x\trans(t)(A\trans + A)x(t)$, it follows that $\mteglemma{A}$ is
a sufficient condition for $\mteg=0$.
Necessity can be shown by noting that $(A\trans+A)\nleq0$ implies the existence of an initial perturbation $x(t_o)=x_o$ that yields $\dot{E}(t_o)>0$; 
thus, $E(t)>E(t_o)$ for some time $t>t_o$, and so $\mteg>0$.
Therefore, $(A\trans+A)\le0$ is a necessary condition for~$\mteg=0$.
%
%
%

%
The maximum transient energy growth can be reduced or suppressed by altering the 
system response characteristics via appropriate actuation~$u(t)$.
%
%
%
In principle, a full-state feedback control law ${u(t)=-Kx(t)}$ 
can be used to achieve various control objectives, including optimal 
regulation by means of a linear quadratic regulator~(LQR)~\cite{brogan1991}.
Here, the $K\in\Re^{m\times n}$ is the controller feedback gain.
The closed-loop dynamics of the associated stable full-state feedback system will then be
\begin{equation}
\dot{x}(t) = (A-BK)x(t).
\end{equation}
However, in practice,
the full-state~$x(t)$ is typically not directly available for feedback.
Instead, measurements of the system outputs $y(t)$
can be used to estimate the full-state by means of
a stable state estimator of the form,
\begin{equation}
	\dot{\hat{x}}(t) = A\hat{x}(t) + Bu(t) + L\left[y(t)-C\hat{x}(t)\right]
\end{equation}
where $\hat{x}(t)$ is the state estimate.
The observer gain $L\in\Re^{n\times p}$ is chosen to yield desirable estimator performance, including adequate convergence rates and reliability 
in the face of both process and measurement uncertainties, as in the case of the optimal state estimator (i.e.,~the Kalman-Bucy filter).
%
%
%

The well-known \emph{separation principle} establishes 
that a stabilizing full-state feedback controller and a stable state estimator can be designed 
independently of one another, then combined to yield a stabilizing 
dynamic compensator by means of the observer-based feedback law~${u(t)=-K\hat{x}(t)}$. 
To see this, consider the dynamics 
%
of the closed-loop observer-plant system with state~${\cl{x}(t)=(x(t),\hat{x}(t))}\in\Re^{2n}$,
\begin{eqnarray}
 \ddt\left[\begin{array}{c}x(t)\\\hat{x}(t)\end{array}\right] &=&  \underbrace{\left[\begin{array}{cc}A&-BK\\ LC&\quad A-BK-LC\end{array}\right]}_{\mathlarger{\Asp}}\left[\begin{array}{c}x(t)\\ \hat{x}(t)\end{array}\right].
 \label{eq:spdyn}
\end{eqnarray}
%
%
The observer-plant system in~\eqref{eq:spdyn} can be brought into a form that replaces the estimated state $\hat{x}(t)$ by the estimation error $e(t)=x(t)-\hat{x}(t)$ via similarity transformation~\cite{brogan1991},
\begin{eqnarray}
 \ddt\left[\begin{array}{c}x(t)\\e(t)\end{array}\right] &=&  \left[\begin{array}{cc}A-BK&BK\\ 0&A-LC\end{array}\right]\left[\begin{array}{c}x(t)\\ e(t)\end{array}\right].
 \label{eq:sepprinc}
\end{eqnarray}
Since the closed-loop operator in~\eqref{eq:sepprinc} appears in block-triangular form, 
the eigenvalues of the observer-based feedback system are simply the union of 
the eigenvalues of the full-state feedback system $(A-BK)$ and the estimator 
$(A-LC)$; indeed, since \eqref{eq:spdyn} and \eqref{eq:sepprinc} are 
related by similarity transformation, this establishes the well-known separation principle.
In the remainder, the term \emph{observer-based feedback} will be used to refer to a closed-loop
system as in~\eqref{eq:spdyn}, formed by means of the separation principle.

Although the separation principle provides guarantees 
on closed-loop \emph{stability}, it does not provide any guarantees 
on closed-loop \emph{performance}---an important point that is often overlooked in the context of transition control and transient energy growth suppression.
%
Consider the closed-loop energy $\cl{E}(t)=\cl{x}\trans(t)\cl{Q}\cl{x}(t)$ of the observer-plant system,
with maximum transient energy growth~$\mtegsp$.
Here, the first $n\times n$ sub-block of $\cl{Q}$ will be $Q$, to be consistent with the original state 
energy defined in~\eqref{eq:energy}.
However, as with~\eqref{eq:energy} and without loss of generality, we let $\cl{Q}=I$ in the remainder.
From the \mlemma, 
the closed-loop operator $\Asp$ 
will exhibit $\mtegsp=0$ if and only if ${\mteglemma{\Asp}}$.
Considering this more closely,
\begin{equation}
\Asp\trans + \Asp = \left[\begin{array}{cc}A\trans+A&(LC)\trans-BK\\ LC-(BK)\trans&\quad (A-BK-LC)\trans + (A-BK-LC) \end{array}\right],
\label{eq:aspaspt}
\end{equation}
reveals that $\mtegsp$ will depend on the plant 
$(A,B,C)$ as well as the compensator gains $K$ and~$L$.
Further, since $(A\trans+A)$ is a principal sub-matrix of $(\Asp\trans+\Asp)$, it follows that $\mteglemma{A}$ is a necessary condition for $\mteglemma{\Asp}$~\cite{prussing1986};
thus, again by the \mlemma, $\mtegol=0$
is a necessary condition for $\mtegsp=0$. 
%
Hence, in the context of transient energy growth suppression and 
transition control, for which the open-loop dynamics exhibit $\mtegol>0$, using the separation principle 
will limit closed-loop performance by guaranteeing that $\mtegsp>0$ as well.
%
%
Note that $G=0$---or, equivalently~$\mteglemma{A}$---is a necessary condition for $\mtegsp=0$, 
but not a sufficient condition;
\emph{all} principal sub-matrices must be considered to establish $\mteglemma{\Asp}$~\cite{prussing1986}. 
As such, even if $G=0$, $\mtegsp$ will also depend on the 
particular system~$(A,B,C)$ and the choices of $K$ and $L$, in general.
%
%

Since the maximum transient energy growth~$\mtegsp$ corresponds to the ``cyber-physical'' 
state $\cl{x}=(x,\hat{x})$ of the observer-plant system, it appears
possible that only one of either $x(t)$ or $\hat{x}(t)$ is 
contributing to the transient energy growth.
Thus, we now establish the influence of the separation principle on 
the maximum transient energy growth $\mtegspx$ of the physical plant and  $\mtegspxhat$ of the observer.
The maximum transient energy growth $\mtegspx$ for the physical plant under 
control via the separation principle can be determined by first considering 
a modified energy $\cl{E}_\epsilon(t)=\cl{x}_\epsilon\trans(t)\cl{x}_\epsilon(t)$, 
where $\cl{x}_\epsilon=W_\epsilon \cl{x}$, ${W_\epsilon=\diag(I_n,\epsilon I_n)}$, $I_n$ is the $n\times n$ identity matrix, and $\epsilon>0$ is a scalar.
Then, noting that $\lim_{\epsilon\rightarrow0}\cl{E}_\epsilon(t) = E(t)$ and that
$\Asp$ and $W_\epsilon=W_\epsilon\trans$ commute, it follows from the \mlemma\ that $\mtegspx=0$ if and only if ${\lim_{\epsilon\rightarrow0}W_\epsilon(\Asp\trans+\Asp)W_\epsilon}\le0$.
As before, $\mteglemma{A}$ is a necessary condition for $W_\epsilon(\Asp\trans+\Asp)W_\epsilon\le0$, so we conclude that $\mteglemma{A}$---and, therefore $\mtegol=0$---is a necessary condition for $\mtegspx=0$.
That is, if $\mtegol>0$, then it is guaranteed that $\mtegspx>0$ as well.
The same can be said about the maximum transient energy growth $\mtegspxhat$ 
for the estimator, which corresponds to a case of observer peaking.
To show this, consider the modified energy $\cl{E}_\epsilon(t)$ defined using the coordinate transformation $W_\epsilon = \diag(\epsilon I_n,I_n)$, then proceed as before.

In summary, $\mteglemma{A}$---and therefore, $\mtegol=0$---is a necessary condition for $\mtegsp=0$, $\mtegspx=0$, and $\mtegspxhat=0$.
Thus, if the uncontrolled plant exhibits non-trivial maximum transient energy growth (i.e.,~$\mteg>0$), then it is guaranteed 
that the separation principle will result in non-trivial maximum transient energy growth in closed-loop for 
the physical plant (i.e.,~$\mtegspx>0$), the estimator (i.e.,~$\mtegspxhat>0$), and the cyber-physical 
observer-plant system~(i.e.,~$\mtegsp>0$).
%
%
%
%
%
%
%
The performance limitations identified here apply only to dynamic output feedback laws synthesized using the separation principle---i.e.,~observer-based feedback as in~\eqref{eq:spdyn}.
Not all output feedback control laws are necessarily subject to
these same limitations. 
%
%
As we will see in the illustrative example that follows,
even when a static output feedback law can be determined
to achieve zero maximum transient energy growth---a necessary condition for guaranteeing the existence of a dynamic output feedback compensator that can achieve the same~\cite{whidborne2007}---the separation principle will not (and cannot) yield a closed-loop system that achieves the same if the uncontrolled plant exhibits $\mteg>0$.

\section{Illustrative example: A simple non-normal system}
\label{sec:example}
\begin{figure}[h!]
\subfloat[$R=2<R^*$\quad $(G=0)$]{\includegraphics[width=0.5\textwidth]{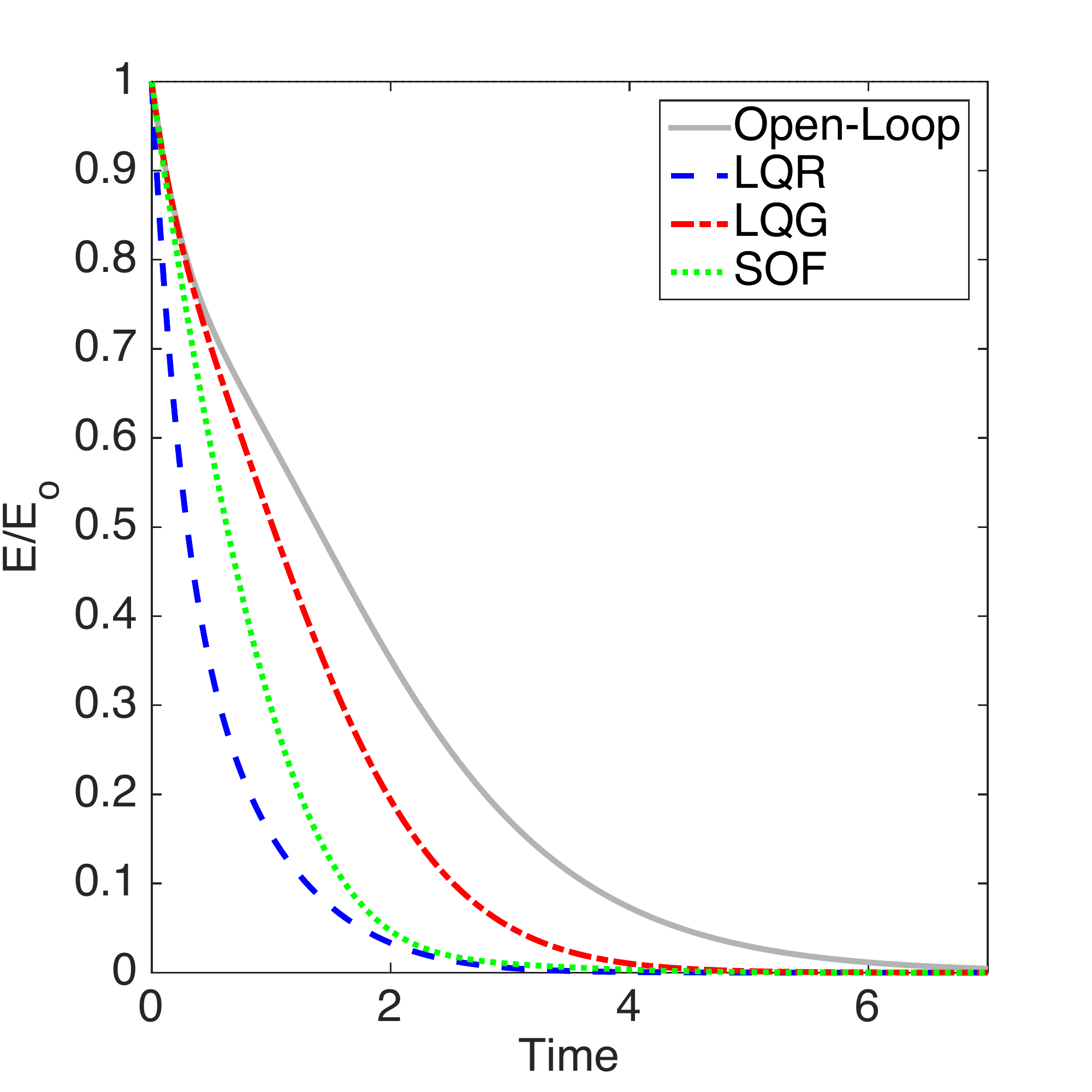}}\hfill
\subfloat[$R=3>R^*$\quad $(G>0)$]{\includegraphics[width=0.5\textwidth]{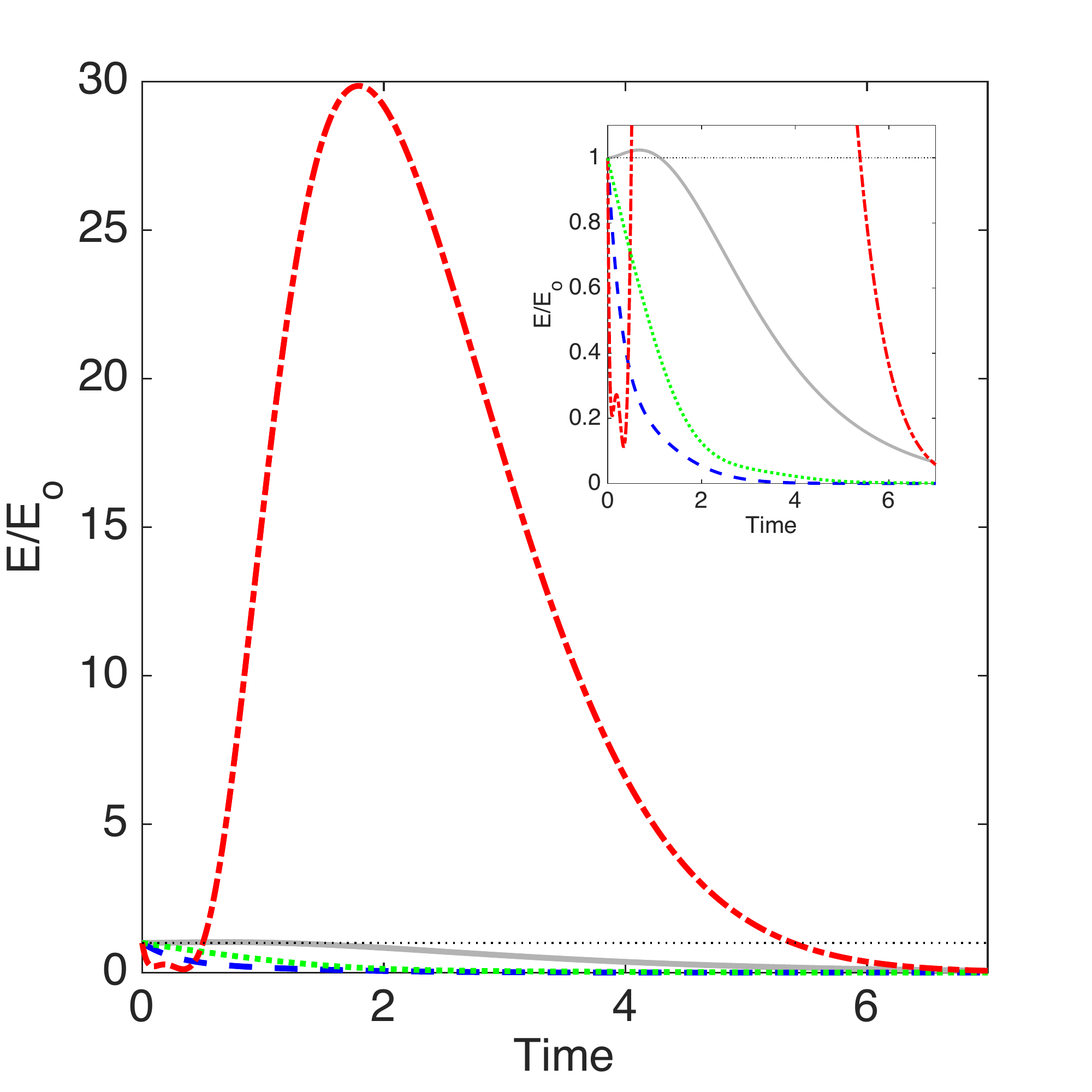}}
\caption{\msh{Comparison of worst-case responses for the controlled system in~\eqref{eq:twobytwo}, for which ${R^*=2\sqrt{2}}$.}}
\label{fig:CLenergy}
\end{figure}

Consider the system,
%
\begin{equation}
\begin{split}
\dot{x}(t) &= \left[\begin{array}{cc}-1/\mathrm{R}&0\\1&-2/\mathrm{R}\end{array}\right]x(t) + \left[\begin{array}{cc}1\\0\end{array}\right]u(t)\\
y(t) &= \left[\begin{array}{cc}0&1 \end{array}\right]x(t)
\end{split}
\label{eq:twobytwo}
\end{equation}
where $R>0$ is a scalar parameter.
Variants of this system are often used to demonstrate 
the role of non-normality in giving rise to transient energy growth~\cite{trefethenScience1993,schmidBook,whidborne2007,hematiAIAA2017}.
The origin is asymptotically stable, but the system 
exhibits $\mteg>0$ for $R>R^*=2\sqrt{2}$. 
In Figure~\ref{fig:CLenergy}, we compare the worst-case transient energy 
response $E(t)$ with $R=2<R^*$ and $R=3>R^*$ for the uncontrolled plant with 
the corresponding worst-case closed-loop system response from each of three different 
controller synthesis techniques: LQR, LQG, and static output feedback~(SOF).
%
%
In all cases, the optimal disturbance for the worst-case response is computed 
by means of the iterative Algorithm~3.1 of Whidborne and Amar~\cite{whidborne2011}.
In the case of LQG control, the worst-case response corresponds to an optimal disturbance on the full observer-plant state $\cl{x}=(x,\hat{x})$.
In instances for which the maximum transient energy growth is zero,
there is no optimal disturbance and so the initial condition is set to either $x(t_o)=(1,0)$ or $\cl{x}(t_o)=(1,0,0,0)$.
%
All energy responses in Figure~\ref{fig:CLenergy} correspond to the energy of the physical plant $E(t)$ normalized by the initial energy $E(t_o) = E_o$.

The LQR design here yields a full-state feedback law that minimizes the performance index,
\begin{equation}
J = \int_{0}^\infty x\trans(t) x(t) + u\trans(t) u(t)\id t
\label{eq:lqrcost}
\end{equation}
subject to the linear dynamic constraint $\dot{x}(t) = Ax(t)+Bu(t)$.
%
%
In the present study, the LQR control gain is computed directly via the 
Matlab command \texttt{lqr}. 
%
%
%
%
%
The response under the LQR full-state feedback control yields $\mteg=0$ 
for both $R=2$ and $R=3$, as seen in Figure~\ref{fig:CLenergy}.
Note that LQR control is not guaranteed to yield zero transient energy growth in general; rather, for this example, the objective function in~\eqref{eq:lqrcost} was specifically selected after 
tuning to achieve zero maximum transient energy growth for the closed-loop response 
for both $R=2$ and $R=3$.

The separation principle is invoked for LQG optimal controller synthesis.
The same LQR control gains determined by~\eqref{eq:lqrcost} are used in 
an observer-based feedback capacity (i.e.,~$u(t)=-K\hat{x}(t)$).
The observer gain $L$ is computed via the Matlab command \texttt{lqe}, which computes a solution to the optimal estimation problem, 
which is dual to the optimal control problem.
Here, the observer gain $L$ was selected by tuning the estimator 
objective function to reduce $\mtegspx$ associated with 
the closed-loop energy response.
Recall, that $\mtegspx$---and the associated optimal disturbance---can be computed by considering ${\lim_{\epsilon\rightarrow0}\cl{E}_\epsilon(t) = E(t)}$.
As expected, in the case of $R=3>R^*$, the LQG controller yields $\mtegspx>0$; whereas, the same LQG controller synthesis approach applied to the case of $R=2<R^*$ yields $\mtegspx=0$.
Further, despite tuning to reduce $\mtegspx$, 
the physical energy $E(t)$ grows to approximately $30$ times its initial value.
In fact, $\mtegspx\approx30\mteg$, meaning the LQG controller degrades the
transient energy growth performance relative to the uncontrolled (open-loop) response.

Lastly, we consider the static output feedback~(SOF) control law~$u(t)=-y(t)$.
By the~\mlemma, the resulting closed-loop operator $(A-BC)$ is guaranteed to exhibit zero maximum transient energy growth for all values of~$R>0$.
Note that the SOF controller considered here is one of 
a family of SOF controllers that can achieve zero transient energy growth~\cite{whidborne2007}.

The results of this simple example illustrate the limitations of the separation principle that were proved in Section~\ref{sec:proofs}.
These results further suggest that controllers synthesized by the separation principle---LQG or otherwise---can be
unreliable for transient energy growth suppression and transition control.
Although the separation principle can greatly simplify controller
synthesis in many instances, the performance consequences of the
 separation principle must also be taken into consideration. 
%
As shown in this example, alternative output feedback approaches 
are not necessarily subject to the same performance limitations and 
should be considered as well as---if not instead of---observer-based feedback.

\section{Discussion and Conclusions}
In this study, we have proven that if the maximum transient energy 
growth for an uncontrolled plant is non-trivial (i.e.,~$\mteg>0$), 
then controller synthesis by the separation principle 
will necessarily result in non-trivial maximum transient energy 
growth in closed-loop (i.e.,~$\mtegsp>0$, $\mtegspx>0$, 
and $\mtegspxhat>0$).
These results were established by invoking the \mlemma\ and properties of negative semi-definite operators 
to show that $A\trans+A\le0$---and thus $G=0$---is a necessary condition for $\mtegsp=0$, $\mtegspx=0$, and $\mtegspxhat=0$.
This result highlights a fundamental performance limitation 
of observer-based feedback control in the 
context of transient energy growth suppression and transition control.
As illustrated in the example of Section~\ref{sec:example}, performance under full-state feedback 
is not necessarily a reliable indicator for closed-loop performance under observer-based feedback.
Indeed, no amount of controller and estimator tuning can overcome the 
performance limitations of observer-based feedback.
If closed-loop transient energy growth is of primary importance, then alternative 
approaches for output feedback control may yield better performance in this regard 
and should also be considered as candidate control approaches.
Further, the result for $\mtegspxhat$ establishes that the 
separation principle will lead to observer peaking from 
some initial conditions in closed-loop whenever $A\trans+A\nleq0$ for the open-loop 
system. 
\msh{Thus, although observer peaking is commonly rooted out as a limiting factor on performance in observer-based feedback control,} this result shows that observer peaking and 
the associated transient 
energy growth of the physical plant can actually be seen as direct 
consequences of using the separation principle to control a system 
for which $\mteg>0$, or equivalently $A\trans+A\nleq0$.

\msh{The results presented here can be generalized further.
Note that all stabilizing controllers can be constructed by means of a $Q$-parameterization, in which a separation-principle-based controller/observer structure is combined with a free parameter~$Q(s)$~\cite{boyd}.
Then, since the separation principle yields a dynamic compensator that is strictly proper, it follows that
when $Q(s)$ is strictly proper, the resulting parameterized dynamic compensator will also be strictly proper; in contrast, when $Q(s)$ is semi-proper, the resulting parameterized dynamic compensator will also be semi-proper.
With this in mind, performing an analysis similar to that of 
Section~\ref{sec:proofs} shows that all strictly 
proper stabilizing controllers will result in non-trivial 
maximum transient energy growth whenever $\mteg>0$.
In contrast, it \emph{may} be possible to achieve zero maximum transient 
energy growth in closed-loop when the dynamic compensator is 
semi-proper---indeed, this corresponds to the analogous 
necessary condition for the generalized result. 
For example, static output feedback control constitutes a 
semi-proper control structure, 
which was seen to fully suppress transient energy growth in the 
illustrative example of this study.
Interestingly, the existence of a static output feedback controller that 
achieves zero maximum transient energy growth is a necessary condition for 
the existence of a dynamic compensator that can achieve the same~\cite{whidborne2007}.
Further, as shown in~\cite{whidborne2007}, a $Q$-parameterization can be used to design controllers that 
minimize the maximum transient energy growth.
%
%
}

Retrospectively, the \msh{performance limitations of 
observer-based feedback for transient energy growth suppression 
are not entirely surprising.}
Consider that the closed-loop operator in~\eqref{eq:sepprinc}
is non-normal.
That non-normality is a necessary condition for transient energy growth 
is well-established in the flow control community;  
it is the high degree of non-normality of the linearized Navier-Stokes 
operator that is commonly attributed to transient energy growth in the 
context of sub-critical transition in shear flows~\cite{schmidBook}.
And yet, the non-normality of the closed-loop operator in~\eqref{eq:sepprinc} 
seems to have evaded attention in many flow control studies.
An important ramification of \msh{this} non-normality is that 
even if a full-state feedback law can fully remove non-normality of the physical  plant \msh{(as in~\cite{hematiAIAA2017})}, 
invoking the separation principle to synthesize 
an observer-based feedback law
will inevitably yield a non-normal cyber-physical observer-plant system.
In this case, although the modes of the physical system 
$(A-BK)$ will still be orthogonal, 
the modes of the coupled observer-plant system 
will be oblique, since the \msh{associated} operator is non-normal.
Further, the degree of non-normality will depend 
on the specific plant~$(A,B,C)$ and the particular 
choice of feedback and observer gains, $K$ and $L$, respectively.
Thus, $K$ and $L$ can be tuned to improve performance for a given 
flow control configuration; yet, if $A\trans+A\nleq0$, then $\mtegsp=0, \mtegspx=0$, and $\mtegspxhat=0$ 
will never be achieved.
Further, the performance limitations of 
the separation principle presented here 
hold true regardless of the system terms~$(B,C)$. 
%
Thus, although efforts at optimal \msh{actuator/sensor} placement and selection
 can be useful at reducing the maximum 
transient energy growth, such efforts will never yield 
$\mtegsp=0$, $\mtegspx=0$, or $\mtegspxhat=0$ when $A\trans+A\nleq0$, 
if observer-based feedback is used for controller synthesis.
%
%
%
Thus, to overcome these performance limitations, alternatives to observer-based feedback control structures
need to be considered as candidates for control.

\section{Acknowledgments}
This material is based upon work supported by the Air Force Office of Scientific Research under award number FA9550-17-1-0252, monitored by Dr. Douglas R. Smith.

\nocite{}
\msh{

}
\end{document}